\pdfoutput=1
\documentclass[runningheads]{llncs}
\usepackage{graphicx}
\usepackage{booktabs}
\usepackage{amsmath}
\usepackage{amssymb}
\usepackage{xcolor}
\usepackage{lmodern}          
\usepackage[T1]{fontenc}
\usepackage[utf8]{inputenc}
\usepackage[expansion=false]{microtype}

\begin{document}

\title{Does Graph Structure Earn Its Place in\\Microservice Root-Cause Analysis?}
\subtitle{A Controlled Study on RCAEval, and What the Benchmark Was Really Measuring}

\author{Imad Bulji\'c}
\institute{Polytechnic Faculty, University of Zenica}

\maketitle

\begin{abstract}
Graph neural networks dominate recent work on microservice root-cause
analysis, yet recent results question whether the graph contributes. Those
results compare whole \emph{pipelines}, so when a flat model wins one
cannot tell whether structure is useless or redundant. We run the
comparison they imply on RCAEval: three learned arms with identical features,
optimiser, validation split, early-stopping rule and scoring head, in which a
single term $W_{\text{neigh}}\!\cdot\!(\hat{A}h)$ separates the graph arms.
Across two RCAEval benchmarks, two topology sources and four regimes we find
\textbf{no reliable
graph-specific effect}: in-distribution the graph model leads the conventional flat
model by $0.003$ Avg@5 ($p=0.844$, $n=6$ disjoint folds).

Auditing the pipeline surfaced two benchmark properties that condition any
result on it. RCAEval injects faults into only five
services per system while exposing 12--70 in telemetry, and the headline
metric is Avg@5: a ranker reading \emph{no telemetry at all} places the true
culprit in the top five on $99.7\%$ of held-out incidents, scoring Avg@5
$0.488$. That prior, not the uniform-random
$0.137$, is the honest in-distribution floor, and it collapses to $0.192$
across systems. The second property is a non-uniform column schema that
silently zeroes telemetry for most RE1 cases, which changes how any
reimplementation of the benchmark should be read. We reproduce a published
baseline (BARO), RCAEval's own reference implementation; on the one system
with a clean schema it reaches similar aggregate accuracy to our own ``naive''
heuristic, within $0.004$, under a different scoring rule, and we read the
divergence elsewhere as a schema effect rather than a quality difference.

The audit motivated a new model. PSC-GRCA separates a candidate score into a
system prior, telemetry evidence, and a centred graph residual, and scores each
term separately. On the six fixed stratified folds it reaches mean Avg@5
$0.915$ against $0.864$ for a flat MLP and $0.862$ for a capacity-matched
no-neighbour control, a fold-level paired Wilcoxon $p=0.03125$, the exact
two-sided minimum at $n=6$. Under transfer it reaches $0.747$ against $0.671$
for the MLP. Ablations locate the in-distribution gain in the prior term:
prior-only scores $0.488$, prior-free drops to $0.850$, and the graph residual
alone reaches $0.800$, so structure adds little on its own. The prior-swap
penalty is a guard rather than the mechanism: the swap passes are made
deterministic, and because the evidence and residual branches read no prior
features the penalty is then exactly zero, so the no-swap runs reproduce the
swapped ones. These results are exploratory: three systems, one architecture
family, and a single prior feature.

We close with a twelve-item checklist for this class of study, distilled from
sixty-two defects recorded while producing the results above, each item
motivated by a failure it would have caught.
\end{abstract}

\section{Introduction}
\label{sec:intro}

Microservice root-cause analysis (RCA) has converged on graph-based
modelling. The service call graph is a natural object: faults propagate along
it, and much recent work builds more expressive encoders over it
\cite{microegrcl,chase}.

Three recent results complicate that consensus. A topology-agnostic MLP matched
GNN-based multimodal diagnosis across five datasets, attributing reported
gains to preprocessing rather than graph modelling; a graph-free method
reached $68\%$ Top-1 on RCAEval's 735 failures; and a propagation-aware
re-evaluation dropped eleven state-of-the-art models to $0.21$ average Top@1
\cite{gao2025,pham2026,fang2025}.

These results compare \emph{systems}, not \emph{representations}. A GNN pipeline
and an MLP pipeline differ in encoder, features, training budget and
preprocessing all at once. When the MLP wins, the comparison cannot say whether
structure is useless or only redundant given that pipeline's features.

\paragraph{This paper isolates the variable.}
Access to neighbours is the only thing that changes. Both arms share the
per-service features (including the within-system rank features), the hidden
width, the dropout and the scoring head. They also share the optimiser and its
learning rate, weight decay and gradient clipping, and the validation
fraction, early-stopping rule and listwise ranking loss.

Isolation is easy to claim and easy to get slightly wrong, so we name the arms.
We run \textbf{three} learned arms, not two:

\begin{itemize}
  \item \texttt{gnn}: the graph encoder;
  \item \texttt{gnn\_no\_neigh}: the same encoder with
        $W_{\text{neigh}}\!\cdot\!(\hat{A}h)$ \emph{and its parameters}
        removed and nothing else changed, so the depth, residual backbone,
        width, head and optimiser are identical;
  \item \texttt{mlp}: a conventional flat scorer of the kind the
        graph-sceptical literature uses.
\end{itemize}

Only the first pair differs by exactly one term. The \texttt{gnn}/\texttt{mlp}
pair differs by $5.8\times$ in parameter count and four layers of depth as
well, because a plain MLP is \emph{not} a GNN with message passing deleted.
The literature often reports the second contrast and calls it the first, so we
report both and label which is which.

\paragraph{Contributions.}
\begin{enumerate}
  \item A controlled isolation of graph access on RCAEval under a matched
        protocol: three learned arms that differ by one term, run across two
        benchmarks, two topology sources and four regimes, with no reliable
        graph-specific effect (Sections~\ref{sec:setup}--\ref{sec:localization};
        the RE2 replication on a \emph{real} Kubernetes co-location topology of
        Section~\ref{sec:re2} is a supporting analysis, not a separate claim).
  \item \textbf{Benchmark-audit findings that condition any result on RCAEval}:
        a telemetry-blind culprit prior worth Avg@5 $0.488$ in-distribution
        that collapses to $0.192$ under transfer, the non-uniform column schema
        that silently zeroes telemetry for most RE1 cases, and what a published
        baseline reveals through that schema lens (Sections~\ref{sec:prior}
        and~\ref{sec:baro}).
  \item A twelve-item checklist for this class of study, each item motivated by
        a specific defect it would have caught (Section~\ref{sec:checklist}),
        together with a prior-separated decomposition, PSC-GRCA, that the audit
        motivated: it scores a system prior, a telemetry-evidence term and a
        centred graph residual separately, and we present it as an exploratory
        attribution instrument rather than a performance contribution
        (Section~\ref{sec:psc}).
\end{enumerate}

\section{Related Work}
\label{sec:related}

\paragraph{Graph methods for microservice RCA.}
The service call graph is the standard substrate for this task, and the
literature has largely pursued more expressive encoders over it. MicroEGRCL
builds a graph neural network with an edge-attention mechanism and enhanced
edge features for root-cause localization \cite{microegrcl}. CHASE encodes
traces, logs and system metrics on a causal hypergraph and localises the cause
with attentive heterogeneous message passing \cite{chase}. Eadro couples
anomaly detection with localization on multi-source data, modelling intra- and
inter-service dependencies in one framework \cite{eadro2023}. Beyond
localization, LEMMA-RCA contributes a large multi-modal, multi-domain dataset
that spans microservice and industrial systems and has become a common
evaluation ground \cite{lemma2024}. RCAEval supplies the telemetry benchmark
this paper uses, together with a reference implementation of BARO that we
reproduce \cite{rcaeval2025}.

Our study does not propose another encoder. It asks whether the graph term
earns its place once every other part of the pipeline is held fixed, the
question raised by the graph-sceptical line of work that reports an MLP at
parity with GNN pipelines \cite{gao2025}, a graph-free method above the
RCAEval baselines \cite{pham2026}, and a propagation-aware re-evaluation that
drops eleven models to $0.21$ average Top@1 \cite{fang2025}.

\paragraph{Evaluation practice.}
Comparisons over a handful of folds are noisy, and the choices that produce a
number matter as much as the number. Dem\v{s}ar recommends non-parametric
signed-rank and Friedman tests for classifier comparison and warns against
repeatedly reusing the same test set \cite{demsar2006}; Dietterich reviews
approximate statistical tests for paired classifier comparison and the
conditions under which each has acceptable error \cite{dietterich1998}.
Bouthillier et al.\ decompose benchmark variance into data sampling,
initialisation and hyperparameter choice, and show that single-configuration
comparisons can be misleading \cite{bouthillier2021}. Henderson et al.\
demonstrate the same for deep reinforcement learning, where runs that differ
only in seed can invert a ranking \cite{henderson2018}. Sculley et al.\ argue
that empirical rigor has not kept pace with empirical output and call for
stronger reporting norms \cite{sculley2018}; Agarwal et al.\ advocate interval
estimates and performance profiles over point estimates at small sample sizes
\cite{agarwal2021}. The fold construction, paired testing and reporting
discipline of this paper follow that line directly, and the twelve-item
checklist in Section~\ref{sec:checklist} is its practical residue.

\section{Experimental Setup}
\label{sec:setup}

\paragraph{Data.}
RCAEval RE1 \cite{rcaeval2025} (Zenodo 14590730, MIT): 375 fault injections across Online
Boutique, Sock Shop and Train Ticket. RE2 Sock Shop adds 90 cases carrying
\texttt{pod-node-*.csv}, the actual Kubernetes placement, so co-location is
exact rather than inferred.

\paragraph{Task and metrics.}
Given a telemetry window around a fault, rank the system's services so the
injected one appears as high as possible. We report AC@$k$ and
Avg@5 $=\frac{1}{5}\sum_{k=1}^{5}\text{AC@}k$, as the benchmark defines them.

\paragraph{A schema trap worth naming.}
RE1's column schema is not uniform. Online Boutique exposes
\texttt{\{svc\}\_latency}/\texttt{\_load}; delay and loss cases expose
\texttt{latency-50}, \texttt{latency-90}, \texttt{workload}; Sock Shop and
Train Ticket expose raw cAdvisor counters. A single literal lookup silently
returns nothing for 250 of 375 cases, yielding zero-filled telemetry and empty
graphs \emph{while appearing to work}. Any reimplementation meets this.

\begin{figure}[t]
\centering
\includegraphics[width=0.86\textwidth]{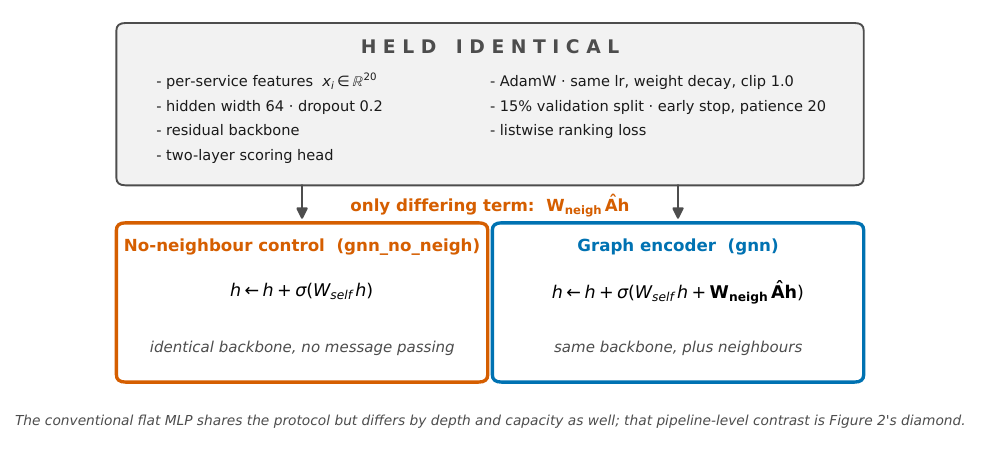}
\caption{The isolation. Everything in the upper box is shared; the two arms
differ by the single bolded term, $W_{\text{neigh}}(\hat{A}h)$. The
pipeline-level comparison the literature reports, a graph encoder against a
flat MLP, differs by depth and capacity as well and is the diamond in
Figure~\ref{fig:forest}.}
\label{fig:isolation}
\end{figure}

\paragraph{Models and matched protocol.}
Both encoders use hidden width 64, dropout 0.2, an identical two-layer head,
AdamW at the same learning rate and weight decay, gradient clipping 1.0, a
$15\%$ validation split and early stopping with patience 20.
Figure~\ref{fig:isolation} states the design.

\paragraph{Folds, and why they are constructed rather than sampled.}
The in-distribution regime uses \textbf{six genuinely disjoint folds},
stratified by (system, fault type), each case assigned to exactly one fold.
An earlier version instead drew an independent $70/30$ resample per seed.
Measured, any two such ``folds'' shared $32\%$ of their test cases and $69\%$
of their train cases, so a paired test across them was not testing what it
claimed. That error produced a significant-looking result which a genuine
partition reduced to noise; we return to it in Section~\ref{sec:localization}.
Cross-system folds are the three held-out systems and were never affected.

\paragraph{Statistics, and an honest limit.}
Paired Wilcoxon signed-rank tests over per-fold values, with bootstrap
intervals. Non-parametric paired tests are the safe choice at these sample
sizes, where the assumptions behind a $t$-test are not met
\cite{demsar2006,dietterich1998}, and repeated measurements of one fold are
averaged to a single value before any test, so a fold never enters the same
comparison twice. The independent unit for transfer is the held-out system,
and RCAEval has three: at $n=3$ no $p$ below $0.25$ is reachable
\emph{whatever} the effect size, so significance testing is underpowered
there by construction. We lead on effect sizes and per-fold values
accordingly. The in-distribution regime reaches $n=6$, where the exact
two-sided Wilcoxon minimum is $p=0.03125$ and nothing smaller is attainable
whatever the effect size. Because a single-run comparison can invert, we report
per-fold values alongside every aggregate \cite{bouthillier2021,henderson2018}.
The same six folds, the same controls and the same early-stopping rule are used
by the prior-separated model of Section~\ref{sec:psc}, so its contrasts are
paired with the arms above; reusing a fixed partition for a second paired
comparison is not the resample error of Section~\ref{sec:localization}, because
no case appears in two folds.

\section{Localization: Structure Has No Reliable Effect}
\label{sec:localization}

\begin{table}[t]
\centering
\caption{RE1 localization, Avg@5 with $95\%$ bootstrap intervals resampled at
the fold. Six disjoint folds in-distribution; three held-out systems
cross-system.}
\label{tab:e7}
\begin{tabular}{lccc}
\toprule
Method & In-distribution & Cross-system & Drop \\
\midrule
Random                                   & 0.137 [0.125, 0.148] & 0.146 [0.044, 0.260] & n/a \\
\textbf{Culprit prior} (no telemetry)    & 0.488 [0.469, 0.504] & 0.192 [0.000, 0.560] & $-0.296$ \\
Max-deviation heuristic                  & 0.797 [0.781, 0.813] & 0.797 [0.627, 0.888] & $+0.000$ \\
MLP (no graph)                           & 0.864 [0.848, 0.880] & 0.671 [0.609, 0.706] & $-0.193$ \\
Same encoder, $-$ neighbour term         & 0.862 [0.847, 0.881] & 0.512 [0.308, 0.692] & $-0.350$ \\
Graph encoder (\texttt{gnn})                             & \textbf{0.867} [0.852, 0.882] & 0.613 [0.459, 0.695] & $-0.254$ \\
\bottomrule
\end{tabular}
\end{table}

Table~\ref{tab:e7} gives the result. \textbf{Structure has no reliable effect
in either regime.} The graph model leads by $0.003$ Avg@5 in-distribution and
trails by $0.058$ cross-system; neither survives a fold-level paired test
($p=0.844$ at a genuine $n=6$; $p=0.25$, the unreachable floor, at $n=3$).

\paragraph{Decomposing the in-distribution margin.}
Splitting \texttt{gnn}$-$\texttt{mlp} into a structure term and a capacity
term:

\begin{center}
\begin{tabular}{llcc}
\toprule
Contrast & Varies & Avg@5 & $p$ \\
\midrule
\texttt{gnn} $-$ \texttt{gnn\_no\_neigh} & message passing alone   & $+0.005$ & 0.313 \\
\texttt{gnn\_no\_neigh} $-$ \texttt{mlp} & depth and capacity alone & $-0.001$ & 1.000 \\
\texttt{gnn} $-$ \texttt{mlp}            & both together            & $+0.003$ & 0.844 \\
\bottomrule
\end{tabular}
\end{center}

Nothing is significant and nothing is directionally stable; per fold the
pipeline-level term is
$[+0.022, +0.013, +0.001, -0.003, -0.012, -0.001]$.

\paragraph{A resample is not a partition.}
An analysis that drew an independent $70/30$ resample per seed appeared to give
a clean, significant result attributed to encoder capacity. The resamples
overlapped, so the paired test was not testing what it claimed, and the effect
does not survive the six-fold partition reported here. A significance test is
only as strong as the partition beneath it.

\paragraph{Cross-system, the decomposition points the other way.}
There the structure term is $+0.101$ and the capacity term $-0.159$ (both
$p=0.25$): the deeper encoder is a \emph{liability} on an unseen system and
message passing recovers part of what it costs. We decline to claim it. The
fold is the held-out system, so $n=3$ is unimprovable, and adopting the graph
arm still costs accuracy overall: the plain MLP beats the full graph model
cross-system. A structure term that buys back a deficit its own architecture
introduced is evidence of entanglement, not of structure helping.

\begin{figure}[t]
\centering
\includegraphics[width=\textwidth]{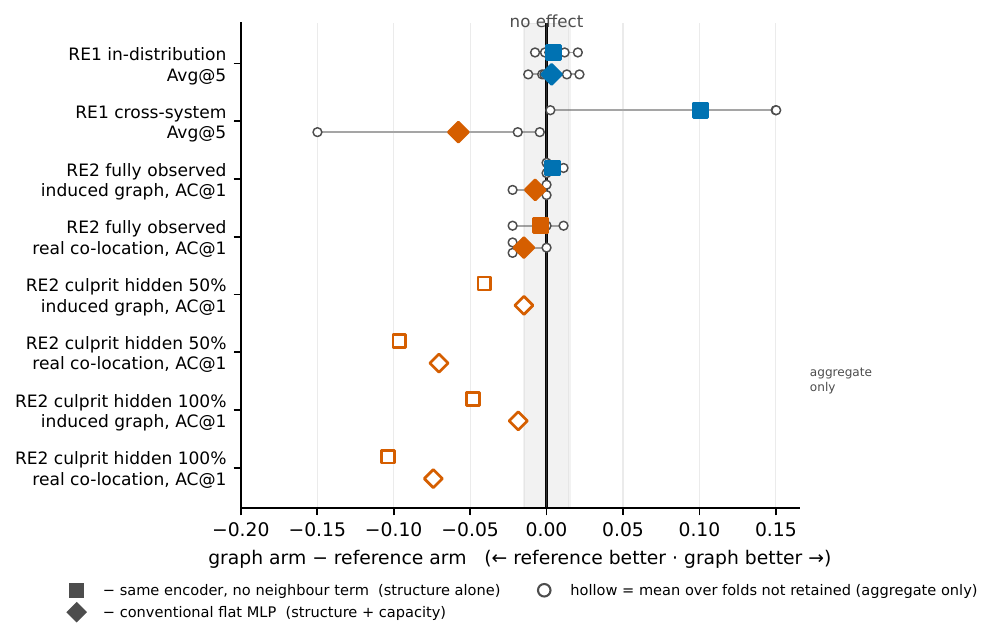}
\caption{Every matched contrast on one axis. \textbf{Squares} compare the
graph arm against the same encoder minus the neighbour term (\emph{structure
alone}); \textbf{diamonds} against a conventional flat MLP (\emph{structure
and capacity}), which is what the literature reports. Filled markers are means
over paired folds, hollow circles the individual folds; coincident fold values
are stacked vertically so multiplicity stays visible. The bottom four rows hold
aggregate scores only, with no per-fold breakdown in the manifest, and are
marked \emph{aggregate only} at the right. Sixteen contrasts, signs in both
directions, nothing surviving as a structure effect.}
\label{fig:forest}
\end{figure}

Figure~\ref{fig:forest} summarizes the complete set of controlled contrasts and
their fold-level uncertainty.

\section{What the Benchmark Was Really Measuring}
\label{sec:prior}

\subsection{A telemetry-blind ranker scores 0.488}

RCAEval injects faults into a small fixed subset of services: \textbf{five per
system} on RE1, against 12, 23 and 70 services actually emitting telemetry,
and five of fifteen on RE2. The headline metric is Avg@5.

So a ranker that knows only \emph{which service names were injected during
training}, and never touches a feature vector, places the true culprit in the
top five almost always. Measured in-distribution: \textbf{Avg@5 $0.488$,
AC@5 $0.997$}. On $99.7\%$ of held-out incidents the true culprit is inside
the top five, from a predictor that cannot see the incident.

That, not uniform-random $0.137$, is the honest in-distribution floor on this
benchmark. Read against it, the best learned arm's margin is $0.379$, not
$0.730$. A fault-injection campaign has to choose targets, so this is a
property of the benchmark rather than a bug in it, but it conditions every
Avg@5 ever reported on RE1.

\subsection{A candidate mechanism for the transfer collapse}

Cross-system the prior collapses to Avg@5 $0.192$ ($0.560$ / $0.016$ / $0.000$
per held-out system), because service names do not transfer. It loses $0.296$
under transfer, \emph{more than either learned model} (MLP $0.193$, GNN
$0.254$), while the max-deviation heuristic loses nothing at all
($0.797 \rightarrow 0.797$).

A model that had absorbed only the prior would lose about as much as the
prior; one that had learned transferable detection would lose about as little
as the heuristic. Both learned models sit between those poles, closer to the
prior. Part of what they learn in-distribution may be \emph{which services
this benchmark injects into}, a component that largely stops paying on an
unseen system.

We state that as a hypothesis, not a measurement. The decomposition bounds the
memorisation component rather than measuring it, and $n=3$ cannot separate the
parts. The prior does not fall to zero under transfer either; it retains
$0.192$, which is why the hypothesis is not an identity, and it is more
specific than an appeal to feature scales.

\paragraph{A hygiene note.}
On Sock Shop and Train Ticket the service list derived from column names
includes node-exporter targets (e.g.\ \texttt{192-168-15-8-9100}) alongside
real services. This inflates apparent ranking difficulty rather than the
prior, and is worth knowing for anyone reporting AC@$k$ on RE1.

\section{A Prior-Separated Decomposition: Where the Score Comes From}
\label{sec:psc}

The audit changes the question a graph model should answer. On RE1 a
telemetry-blind prior already reaches Avg@5 $0.488$, and the controlled study
shows that adding a graph term to a matched encoder moves the score by
$0.005$. A model that reports a single high score is therefore uninformative
about its own behaviour: it may be reading the prior rather than the incident.
We build the separation into the architecture so that each part can be scored
on its own.

\paragraph{Decomposition.}
PSC-GRCA, for prior-separated counterfactual graph RCA, scores each candidate
service $i$ as
\begin{equation}
s_i \;=\; p_i \;+\; e_i \;+\; r_i ,
\label{eq:psc-score}
\end{equation}
where $p_i$ is a system prior read from the training-time distribution of
injected services, $e_i$ is telemetry evidence from the service's own feature
vector, and $r_i$ is a graph residual from a two-layer relational encoder over
the induced adjacency. All three terms are unit-width heads on the same
optimiser.

\paragraph{Centred residual.}
The residual is constrained to have zero mean over the candidate set,
\begin{equation}
r_i \;\leftarrow\; r_i \;-\; \frac{1}{|C|}\sum_{j \in C} r_j ,
\label{eq:psc-centre}
\end{equation}
so $r$ cannot absorb the constant offset that a system-level prior already
supplies. Only the shape of the graph signal survives, which is what a
structure claim should be about.

\paragraph{Prior-swap contract.}
Training can add a penalty asking the evidence and residual terms to be
invariant when the prior is permuted,
\begin{equation}
\mathcal{L}_{\text{swap}} \;=\; \bigl\lVert (e_i + r_i) - (e'_i + r'_i) \bigr\rVert_2^2 ,
\label{eq:psc-swap}
\end{equation}
Equation~\ref{eq:psc-swap} is the prior-swap penalty: primes denote a pass with the prior
vector rolled by one position, both passes run with dropout disabled so the term measures prior invariance rather
than dropout consistency. Because the swap passes are deterministic and the
evidence and residual branches take no prior input, the penalty is exactly zero
here and contributes no gradient; we keep it as a guard so that a later variant
which couples those branches to the prior cannot do so silently. The model is
trained end to end with cross-entropy on $s$, plus
$0.05\,\mathcal{L}_{\text{swap}}$ when the penalty is enabled. The no-swap arm
omits that term and changes nothing else; since the term is exactly zero the
two arms reproduce each other at the same seed, so the no-swap runs are a check
that the reported numbers do not depend on the penalty, not evidence about
which term carries the result.

\paragraph{Protocol.}
The model reuses the controlled study's six fixed stratified folds and three
held-out systems, with the same $15\%$ validation split taken from the tail of
each training fold, the same optimiser (AdamW, learning rate $10^{-3}$, weight
decay $10^{-5}$), the same gradient clipping ($1.0$) and the same
early-stopping rule: patience 20 on the validation listwise ranking loss, at
most 200 epochs. The model is trained on the full score $s = p + e + r$. The
ablation arms are \emph{scoring} evaluations of that same trained model, one
term of Equation~\ref{eq:psc-score} selected at scoring time, not separately
trained runs, so a gap between arms cannot be an optimisation difference
between them.

\begin{table}[t]
\centering
\caption{PSC-GRCA ablations, Avg@5. \emph{Stratified} averages the six fixed
folds; \emph{cross-system} averages the three held-out systems. One model is
trained per fold under the protocol above, always on the full score; each arm
scores that same model on the term named in its row.}
\label{tab:psc}
\begin{tabular}{lcc}
\toprule
Arm & Stratified & Cross-system \\
\midrule
Full ($p+e+r$)        & \textbf{0.915} & \textbf{0.747} \\
Prior-free ($e+r$)    & 0.850 & 0.747 \\
Prior-only ($p$)      & 0.488 & 0.079 \\
Evidence-only ($e$)   & 0.853 & 0.673 \\
Graph-only ($r$)      & 0.800 & 0.692 \\
\bottomrule
\end{tabular}
\end{table}

\paragraph{The in-distribution gain is mostly the prior.}
Table~\ref{tab:psc} gives the ablations. On the six fixed stratified folds the
full model reaches mean Avg@5 $0.915$ against $0.864$ for the flat MLP and
$0.862$ for the capacity-matched no-neighbour control of
Section~\ref{sec:localization}. The paired per-fold differences against both
controls are positive in every fold, and a two-sided Wilcoxon signed-rank test
over the six folds returns $p=0.03125$, the exact two-sided minimum at $n=6$.
It is the only contrast reported here that clears the conventional threshold,
and it is a contrast between the prior-separated model and the flat controls,
not a graph-structure effect: prior-free falls to $0.850$, below the flat
baseline, while prior-only scores $0.488$, so the full model's margin over that
baseline comes from the prior term. The graph residual alone reaches $0.800$
and adds nothing on top of the evidence term ($0.853$). We read the result as
evidence that a learned model can exploit the benchmark prior and that
separating it makes the attribution visible, not as evidence that graph
structure is necessary. Read that way the gain locates in the prior term, the
graph residual adds nothing over the evidence arm, and the prior-swap penalty
is exactly zero in this architecture, so PSC-GRCA is an instrument for
attribution rather than a performance contribution.

\paragraph{Cross-system.}
Under transfer the full model reaches $0.747$ against $0.671$ for the flat MLP.
The prior-only arm collapses to $0.079$, consistent with the collapse of the
culprit prior reported in Section~\ref{sec:prior}, and the prior-free arm also
scores $0.747$. That equality has a mechanical explanation, and it matters for
reading the table. The prior feature is a training-frequency count over service
names, and on $374$ of the $375$ held-out cases every candidate carries a zero
count, because the held-out system's service names never appear in the training
systems. A constant term cannot change a ranking, so the prior shifts every
score by the same amount and the two arms coincide to the reported precision;
the single exceptional case does not move the aggregate at three decimals. The
prior-only number should be read the same way. With a constant prior the
ranking is decided entirely by the scorer's tie order (the score vector sorted
by the default \texttt{numpy} argsort and reversed, in
\texttt{pilot\_psc\_grca.py}), so $0.079$ measures that tie
convention rather than transferable prior knowledge. In-distribution the same
arm scores $0.488$ because the counts there are nonzero and informative.
The three cross-system folds cannot support a significance claim, and that
reading is weak for the reason that follows: a separately trained arm, or one
scored under a different objective, cannot attribute a score to a term, and
with every arm scored from one trained model we cannot separate a term that
contributes from a term the fit merely leaves small. A companion check disables
the prior-swap penalty and reruns the folds: cross-system the no-swap arm scores
$0.747$, indistinguishable from the swapped run, and eight independent
stratified seeds span $0.914$ to $0.928$ (mean $0.921$). Because the penalty is
exactly zero once its passes are deterministic, those runs reproduce the
swapped runs at the same seed. The check shows the reported numbers do not
depend on the penalty. It is a check, not proof of the mechanism.

\paragraph{What this is not.}
The study is exploratory. Three systems, one architecture family, and a single
prior feature leave both the transfer claim and the ablation attribution
under-powered. The arms are scoring evaluations of one trained model, so they
describe which term the fitted model uses, not what training without that term
would yield, and a separately trained or differently scored arm could not close
that gap either. RE2 is covered here only by the controlled analyses of
Section~\ref{sec:re2}. The prior-separated model has not been ported to the RE2
loader, which reads the real Kubernetes co-location topology: the localization
loader in this artifact (\texttt{kintsugi/data/localization.py}) reads RE1 only,
and an earlier PSC run labelled RE2 was scored on RE1 folds. We withdraw that
result and make no PSC claim on RE2 until the model is ported. We present
PSC-GRCA as an exploratory attribution instrument that the audit motivated,
not as a performance contribution or a settled method.

\section{What a Published Baseline Reveals About the Schema}
\label{sec:baro}

BARO is RCAEval's own reference implementation~\cite{baro2024}, taken from the
authors' repository and invoked as their benchmark harness invokes it, scored
on the same cases, folds and metric as every other arm.

\begin{table}[t]
\centering
\caption{BARO against our max-deviation heuristic, per system. Column counts
are for one case per system, since raw frame width varies (Online Boutique
51--60, Sock Shop 421--439, Train Ticket 1180--1446); the values and the script
that counts them are recorded in \texttt{output/real/dataset\_facts.json}.}
\label{tab:baro}
\begin{tabular}{lccl}
\toprule
System & BARO & Max-deviation & Raw frame \\
\midrule
Online Boutique & \textbf{0.880} & 0.877 & 51 columns, curated RED metrics \\
Sock Shop       & 0.557 & \textbf{0.888} & 438 columns, raw cAdvisor \\
Train Ticket    & 0.277 & \textbf{0.627} & 1242 columns, raw cAdvisor \\
\midrule
Overall         & 0.571 & 0.797 & \\
\bottomrule
\end{tabular}
\end{table}

Table~\ref{tab:baro} is a published-baseline check, not a leaderboard.

\textbf{On Online Boutique the two reach similar aggregate accuracy: $0.880$
against $0.877$, a difference within $0.004$.} That similarity needs care,
because the two methods do not score the same way. The reference
implementation fits a \texttt{RobustScaler} to each column's pre-fault window,
transforms the post-fault window with it, and ranks columns by the maximum
signed transformed sample~\cite{baroimpl}. Our heuristic standardises the
post-fault window mean by the pre-fault mean and standard deviation, takes the
absolute value, clips it, and keeps the largest over four curated metric
families. Where the reference implementation scores a single sample, ours
averages the post-fault window. The reference scales by median and
interquartile range; ours scales by mean and standard deviation. And while the
reference ranks raw columns, ours ranks curated families aggregated to
services. The
two can land on the same aggregate accuracy on one system while disagreeing
about individual incidents. The RCAEval harness supplies the injection time,
so both run without BARO's online change-point estimation and only the scoring
stage is compared. The agreement is evidence that the heuristic is a
reasonable comparator on this benchmark. It is not evidence that the two rules
are the same rule.

The divergence elsewhere is a \emph{schema} effect, not a quality one, and it
does not make our heuristic better. BARO ranks individual columns; on the
cAdvisor systems those are hundreds to over a thousand raw counters, so any
one spiking counter lifts a service. Ours aggregates each service to four
curated metric families through the alias table that exists only because RE1's
schema is non-uniform. The gap measures what per-service metric curation is
worth here. BARO's $0.571$ should be read as ``BARO under our
column-to-service mapping, without dataset-specific column selection'', not as
a refutation of a published method.

\section{RE2: A Real Topology Does Not Change the Answer}
\label{sec:re2}

RE1 ships no dependency graph, so ours is induced from correlation, which
leaves open that the null reflects a poor graph. RE2 closes that: co-location
is read from actual Kubernetes placement, and it is the mechanistically
appropriate topology, since four of six RE2 fault types are resource
contention propagating through a shared machine.

Fully observed, every arm sits near ceiling (AC@1: MLP $0.978$, no-neighbour
control $0.967$, induced graph $0.970$, real co-location $0.963$) and no
pairwise difference approaches significance at $n=3$.

\paragraph{The blind-spot regime.}
The condition that creates headroom is the realistic inverse: the faulty pod
stops exporting, so the one service you must find is the one you cannot see.
We had hypothesised a structural mechanism: a flat model has no signal, while
a graph can implicate the service through neighbours sharing its node.
\textbf{A single-seed pilot supported this; full replication does not.} The
manifest keeps only aggregate scores at each masking rate, not a per-fold
breakdown, so every number in this paragraph is a point estimate and none of
them carries a significance claim. At $100\%$ masking the capacity-matched
no-neighbour control ($0.104$) sits above the induced graph ($0.056$) and the
real co-location graph ($0.000$). The arm with \emph{no} message passing has
the highest point estimate in the one regime where the hypothesis said it
should be lowest, and the real-topology arm is lowest.

One pattern holds across those aggregate point estimates: at non-total masking
the learned arms sit above the naive heuristic, while at $100\%$ masking the
heuristic and the real-topology arm both fall to zero. That pattern is worth
keeping, and it is not a pattern about graphs. Whether it holds fold by fold
cannot be tested from the retained evidence.

\begin{figure}[t]
\centering
\includegraphics[width=\textwidth]{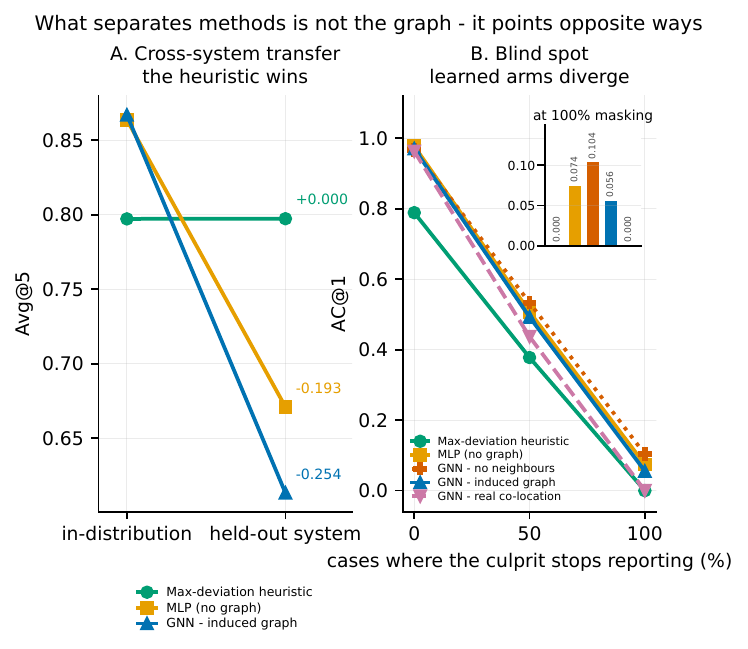}
\caption{The two hard regimes disagree about the same method. \textbf{A}~Under
transfer the heuristic loses essentially nothing while both learned models
lose $0.193$ and $0.254$. \textbf{B}~Under a blind spot the ordering reverses.
Panel~B is aggregate only: the manifest retains no per-fold breakdown at each
masking rate, so its ordering is a point estimate with no significance claim.
Neither panel separates the graph arms from the flat one.}
\label{fig:regimes}
\end{figure}

Figure~\ref{fig:regimes} visualizes the opposing transfer and blind-spot regimes.

\section{A Checklist for Graph-versus-Flat Ablation Studies}
\label{sec:checklist}

The checklist is the practical residue of the study: we recorded sixty-two
defects while producing the results above and distilled twelve checks from
them, each motivated by a specific failure it would have caught. Roughly half
of those failures produced numbers that looked plausible rather than obviously
wrong, and would have survived ordinary code review. The items below name the
failure each check addresses.

\begin{enumerate}\itemsep2pt
  \item \textbf{Audit label leakage at the task level, not the feature level.}
        \emph{Caught: an early graph encoded its own ground truth through a
        task-level leakage path.}
  \item \textbf{Assert that message passing reaches the nodes being scored.}
        \emph{Caught: no inverse relations, so the fault node received zero
        messages; the symptoms are indistinguishable from a clean negative
        result.}
  \item \textbf{Give every compared model the same training protocol, checked.}
        \emph{Caught: a control that ended in a double sigmoid and trained
        $50$ fixed epochs against the graph's $200$-with-early-stopping;
        correcting it reversed the headline.}
  \item \textbf{Never let an oracle answer the question it is meant to test.}
  \item \textbf{Compute significance over independent folds, never repeated
        measurements.} \emph{Caught: recording a deterministic baseline once
        per seed inflated $n$ from 3 to 9 and manufactured $p=0.0117$.}
  \item \textbf{Treat every single-run result as a pilot.}
  \item \textbf{Verify your verification tooling with a negative control,
        every time it changes.} \emph{Caught twice; the second time was the
        first fix reintroducing the bug.}
  \item \textbf{Never accept a between-arms verdict without a real test.}
  \item \textbf{Generate every figure from the evidence files, and make the
        build deterministic.}
  \item \textbf{Write the decision rule down before you have the result.}
  \item \textbf{A resample is not a partition.} Seeding a random resample
        differently per ``fold'' looks like cross-validation and is not.
        \emph{Caught: folds sharing $32\%$ of their test cases produced
        $p=0.031$ for an effect a true partition puts at $p=0.844$.}
  \item \textbf{Check that every baseline and control can actually vary its
        output.} Print each one's distinct-prediction set. \emph{Caught: a
        ``compatibility oracle'' that emitted the same operation for all 375
        incidents, and an ablation that emitted one operation for the entire
        test set. They were reported as two independent controls and were
        identical bit-for-bit.}
\end{enumerate}

A negative result is publishable only if the thing negated was actually built,
replicated, checked against its own tooling, tested rather than eyeballed,
reported under a rule fixed beforehand, computed over folds verified
independent, and compared against controls verified non-degenerate. The list
above is the practical residue of applying those standards here, and it is
what we would ask of any study that reports a null of this kind
\cite{sculley2018,agarwal2021}.

\section{Limitations}
\label{sec:limitations}

\textbf{One published baseline is not a leaderboard.} BARO is metric-only and
training-free; the graph-based methods our motivation argues with are not
reproduced here, and those are the ones whose pipelines would test the
confound directly.
\textbf{One architecture family.} The graph arm is a GraphSAGE-style neighbour
aggregator with a single edge type throughout; confirming the null is not
architecture-specific needs a second.
\textbf{Cross-system transfer is three folds}, which caps what any test can
show.
\textbf{RE2 is a single system}, 90 cases.
\textbf{The blind-spot regime is simulated}; real telemetry loss correlates
with the fault rather than being independent of it.
\textbf{PSC-GRCA is exploratory.} Its transfer estimate rests on the same three
held-out systems, so the cross-system contrast cannot reach significance by
construction. A single architecture family and a single prior feature mean the
ablation cannot separate the prior's contribution from the particular
parameterisation that produced it. The arms are scoring evaluations of one
trained model, so they show which term the fitted model uses rather than what
training without a term would yield, and the no-swap check is a reproducibility
check on the penalty, not proof of the mechanism. RE2 remains covered only by
the controlled analyses of Section~\ref{sec:re2}: the prior-separated model has
not been ported to the RE2 loader, so no PSC claim is made there. The
in-distribution contrast sits at the exact two-sided Wilcoxon minimum for six
folds ($p=0.03125$); no smaller $p$ is attainable at this $n$ whatever the
effect size, so the effect size and the per-fold values are what should be read,
not the significance threshold.
\textbf{Precision, not validity, in some studies.} Some studies pair
each fold with a single model initialisation rather than averaging several, so
fold identity and initialisation are aliased. This does not bias the paired
comparison, since both arms see the same pairing, but it leaves more
initialisation noise in each fold value than necessary, which widens spread
and makes detection \emph{harder}. The direction is conservative, and we
report it rather than spend six times the compute narrowing a null.

\section{Conclusion}

Under matched protocol on RCAEval, across the tested model families, two
benchmarks, two topology sources and the four regimes we test, graph structure
has no reliable effect on microservice root-cause localization. What separates
the arms here is whether the model is learned at all, and which of two opposite
hard conditions is tested, not access to a graph. A one-line deviation
heuristic beats both learned models under cross-system transfer and loses
decisively under a blind spot. That blind-spot turn is aggregate only: the
manifest retains no per-fold breakdown at each masking rate, so it is a point
estimate and carries no significance claim.

What auditing the benchmark revealed may prove the more durable contribution.
Half of the in-distribution Avg@5 on RCAEval RE1 is obtainable by a ranker
that reads no telemetry, because the injection campaign targets five services
per system and the metric is Avg@5. The same prior collapses across systems,
which gives the transfer gap a candidate mechanism and is more specific than an
appeal to feature scales. The non-uniform schema is a second property that
changes how a reimplementation should be read, and the published baseline we
reproduced reaches similar aggregate accuracy to the heuristic we had
introduced as a floor, though under a different scoring rule. None of these are
facts about graph neural networks; all of them condition what a number on this
benchmark means.

The audit also motivated a model that takes the prior seriously. PSC-GRCA
separates the prior from telemetry evidence and a centred graph residual, and
reaches mean Avg@5 $0.915$ on the six fixed stratified folds, the only contrast
there that clears the conventional threshold, against $0.747$ under transfer.
Its ablations locate most of the in-distribution gain in the prior term, so the
honest reading is that separation makes the attribution visible rather than
that structure has been vindicated. The result is exploratory and leaves the
transfer question open. The prior-swap penalty is a guard in this architecture:
the swap passes are deterministic, so the term is exactly zero, and the no-swap
runs confirm only that the reported numbers do not depend on it.

\paragraph{Follow-up direction.}
The next step is a second architecture family and a second prior feature, to
separate the contribution of the separation itself from the particular heads
used here, and a larger set of held-out systems so that transfer can be tested
rather than only estimated. Prior-swapped incidents remain the intended
instrument for a future variant that couples the branches to the prior, where
the swap penalty becomes non-trivial; in the present architecture it is a
guard, not a mechanism.

\paragraph{Artifact.}
Code, the seed manifests behind the main study, a sixty-two-entry defect
register, and the scripts that turn manifests into the tables and figures are
maintained with the research artifact. The checker validates the
\texttt{preprint.tex} claims against those manifests and against two record
files: \texttt{output/real/dataset\_facts.json} holds dataset sizes and the
defect count, and \texttt{output/real/psc\_grca\_summary.json} holds every
PSC-GRCA decimal, computed from the run artifacts rather than transcribed,
because the checker extracts only three-decimal results. The checker refuses any
record key that merely restates the manuscript, so a number cannot be verified
against its own transcript.
Public artifact release is separate from this preprint.

\end{document}